\documentstyle[prl,aps,epsfig]{revtex}
\draft
\begin{document}

\twocolumn[
\hsize\textwidth\columnwidth\hsize\csname@twocolumnfalse\endcsname

\title{Influence of the Kondo effect in a non-Fermi liquid system}
             
\author{Karyn Le Hur}
\address{Laboratoire de Physique des Solides, Universit{\'e} Paris--Sud,
		    B{\^a}t. 510, 91405 Orsay, France}
 \maketitle

\begin{abstract}
The Kondo effect in a Tomonaga-Luttinger liquid ($U<<t$) is studied by using the non-Abelian bosonization. The $q=2k_F$ enhanced spin fluctuations generate a special Kondo effect, for any sign of the exchange coupling $J_K$ ($\left|J_K
\right|<<U$) with the impurity. Then, unlike in Fermi liquids 
($U\rightarrow 0$), the presence of a $2k_F$-polarized screening cloud around 
the impurity favors the occurrence of irrelevant electronic operators with scaling dimension $d=3/2$: the thermodynamics is nearly this of the two-channel Kondo model in a Fermi liquid. The Mott insulating transition does not affect much the ground state, but rescales the power-law dependence of the Kondo temperature on $J_K$.

\end{abstract}

\vfill
 \twocolumn
\vskip.5pc ]
\narrowtext

The antiferromagnetic exchange interaction $J_K$ between a {\it free} electron gas and a localized magnetic impurity gives rise to the so-called Kondo effect\cite{un}. When the temperature reaches the characteristic Kondo energy scale $T_k\propto e^{-1/\rho J_K}$, where $\rho$ is the density of states per spin, the local moment forms a singlet pair with a conduction electron within an energy $T_k$ of the Fermi energy\cite{deux}. In the scaling language, the electrons participating in the formation of this local singlet lie within a characteristic distance $\xi_{imp}\sim v_F/T_k$ where $v_F$ is the Fermi energy. This is a length that is far greater than the lattice spacing and it tends to form
a real ``screening cloud'' around the impurity site. The low-temperature 
behavior seems trivial. The impurity has disappeared from the low-energy 
physics, but certains interactions between electrons are generated in the 
processes of eliminating the impurity spin. The Kondo effect can
be mapped to a one-dimensional model, and  much of the interesting 
behavior comes from {\bf two} leading irrelevant electronic
operators, which have exactly the same scaling dimension $d=2$. Nozi\` eres\cite{trois} argued that they have an universal ratio, so there is only one unknown parameter, the ``Wilson number''. We expect the concerned irrelevant coupling constant to be ${\cal O}(1/T_k)$ by a standard scaling argument. The first order perturbation theory in it, gives a {\it heavy-fermion} behavior defined by a linear specific heat $C_{imp}\sim T/T_k$, a constant susceptibility $\chi_{imp}\sim T/T_k$ and an universal Wilson ratio $R_W=2$.

In three dimensions, electron-electron interactions do not affect the Kondo fixed point, just dress up bare electrons into nearly independent 
quasiparticles. On the other hand, the Landau's Fermi liquid theory breaks down in a one dimensional electron gas, for {\it arbitrarily} weak interactions $U$\cite{quatre}. The effect of electron-electron interactions is particularly strong in one dimension and leads to a non-Fermi liquid state (the so-called Tomonaga-Luttinger (TL) liquid\cite{cinq}). In the presence of repulsive interactions, quasiparticle like excitations are replaced by collective density spin and charge excitations, namely {\it spinons} and {\it holons}. Recently, considerable attention has been focused theoretically on the response of such a system to localized perturbations\cite{six}; it
would be experimentally accessible with narrow single-channel quantum 
wires. Classically, the TL liquid is governed by $q=2k_F$
charge and especially spin density waves; so, we are led to think that the 
Kondo effect in a TL liquid is totally different from 
that in a Fermi liquid. Now, it has been well established that sufficiently 
enhanced $q=2k_F$ spin fluctuations also support a special Kondo effect, for 
ferromagnetic as well as antiferromagnetic Kondo exchanges\cite{sept,huit}. The Kondo 
temperature is expected to yield a power-law dependence on the exchange 
coupling $(\left|J_K\right|<U)$, due to the particular density of states of 
the TL liquid in the vicinity of the fixed point. But, the different used 
weak-coupling analyses\cite{sept,huit} do not allow to give a precise sight of 
the ground state in the regime $U>J_K$. Recent exact results from Conformal 
Field Theory (CFT)\cite{onzei} are available only in the
limit $U<J_K$. In this Letter, we address the following questions: What is the 
fixed point of the Kondo model in such a non-Fermi liquid system for 
$U>J_K$? What are the pseudo-particles in the ``screening cloud''? 

Our starting point is the Hamiltonian:
 \begin{eqnarray}
 \label{zero}
 H&=&-t\sum_{i,\sigma} c^{\dag}_{i,\sigma}c_{i+1,\sigma}+ (h.c) + U\sum_{i,\sigma}n^c_{i,\sigma}n^c_{i,-\sigma}\\ \nonumber
 &+&J_K\ c^{\dag}_{i=0,\alpha}(\vec{\sigma}^{\alpha}_{\beta}/2) c_{i=0,\beta}.\vec{\tau}
\end{eqnarray}
where the bare parameters obey $(U,J_K)<<t$. Here, $c^{\dag}_{i,\sigma}$ $(c_{i,\sigma})$ creates (annihilates) an electron of spin $\sigma$ at site i and
$\vec{\tau}$ is a spin $\frac{1}{2}$ operator located at site $i=0$. A small
U-Hubbard interaction between c-electrons has been introduced, and $J_K$ describes the Kondo coupling. To preserve the SU(2) spin symmetry, to
include the effects of the $4k_F$-Umklapps at half-filling, and to
analyze precisely all the irrelevant operators, we use a continuum limit of the
Hamiltonian and we switch over to non-Abelian bosonization 
notations\cite{neuf}.

{\it Non-Abelian notations of the (TL) liquid:} We linearize the dispersion of conduction electrons; the lattice step is fixed to $a=1$. The relativistic fermions $c_{\sigma}(x)$ are separated in left-movers $c_{L\sigma}(x)$ and right-movers $c_{R\sigma}(x)$ on the Fermi-cone. As usual, we introduce the ``normal-ordered'' current operators for the charge and spin degrees of freedom, namely $J_{c,L}=:c^{\dag}_{L\sigma}c_{L\sigma}(x):$ and $\vec{J}_{c,L}=:c^{\dag}_{L\alpha}\frac{{\vec{\sigma}_{\alpha\beta}}}{2}c_{L\beta}:$ and similarly for the right-movers. The charge Hamiltonian is equivalently described in terms of the
massless scalar field $\Phi_c$ and its moment conjugate $\Pi_c$:
\begin{eqnarray}
 H_c&=&\int dx\  \frac{u_{\rho}}{2 K_{\rho}}:{(\partial_x\Phi_c)}^2:+\frac{u_{\rho}K_{\rho}}{2}: {(\Pi_c)}^2:\\ \nonumber
      &+&g_3\exp(i4k_Fx)\cos(\sqrt{8\pi}\Phi_c)
\end{eqnarray}
The coupling $g_3\propto U$ generates the usual $4k_F$-Umklapp process 
{\it only} relevant at half-filling, and the parameters $u_{\rho}$ and 
$K_{\rho}$ which describe the TL liquid are given by:
 \begin{equation}
  u_{\rho}K_{\rho}=v_F \qquad \text{and} \qquad \frac{u_{\rho}}{K_{\rho}}=v_F+2U/\pi
  \end{equation}
The Fermi velocity is $v_F=2t\sin k_F$. The spin sector is described by a k=1 Wess-Zumino-Witten (WZW) Hamiltonian\cite{dix}:
\begin{equation}
 H_s=\frac{2\pi v_F}{3}\int\ dx\ :\vec{J}_{c,L}({x})\vec{J}_{c,L}({x}):+(L\rightarrow R)
\end{equation}
To treat the local Kondo interaction, we need the complete representation
for the conduction spin operator\cite{neuf}:
  \begin{eqnarray}
 \vec{S}_c&\simeq&\vec{J}_{c,L}(x)+\vec{J}_{c,R}(x)\\ \nonumber
 &+&\alpha\exp(i2k_F x)tr(g.\vec{\sigma})(x)\cos(\sqrt{2\pi}\Phi_c)(x)
 \end{eqnarray}
where $\alpha$ is a simple constant. For the following, we do not need explicitly the exact representation of the current operator $\vec{J}_{c,L}(x)$ in 
terms of the spin matrix g. Finally, the two relevant spin couplings come out 
as:
\begin{eqnarray}
&\lambda_2&(\vec{J}_{c,L}(0)+\vec{J}_{c,R}(0))\vec{\tau}\\ \nonumber
&+&\lambda_3 tr(g.\vec{\sigma})(0).\cos(\sqrt{2\pi}\Phi_c)(0)\vec{\tau}
\end{eqnarray}
where $\lambda_{i=2,3}\propto J_K$. The couplings $\lambda_2$ and
$\lambda_3$ respectively generate the {\it forward} and {\it backward} Kondo scattering processes. One aim of this paper is to clarify the fact, that a
heavy-fermion fixed point cannot occur in a TL liquid, so we start with
the bare constraint $J_K<<U$. The effects of the electronic
interactions are crucial in that case.

{\bf Luttinger liquid away from half-filling}: We first re-investigate the weak-coupling regime in which $J_K/\pi v_F<<U/\pi v_F<<1$, preserving
explicitly the SU(2) symmetry. By using the following commutation rules 
$(z=x+v_F.t)$:
\begin{eqnarray}
[{\cal J}_{c,L}^a(z),{\cal J}_{c,L}^b(z')]&=&if^{abc}{\cal J}_{c,L}^c(z)\delta(z-z')\\ \nonumber
&+&\frac{i}{4\pi}\delta^{ab}\delta'(z-z')
\end{eqnarray}
with ${\cal J}_{c,L}^a={J}_{c,L}^a\delta(x)$, and the integral: $\frac{1}{v_F}\int \frac{du}{u}\sim \frac{ln L}{v_F}$ with $u=v_F.t$, we obtain the usual $\lambda_2$-beta function\cite{neuf}:
\begin{equation}
\beta(\lambda_2)=\frac{d\lambda_2}{dln L}=\frac{\lambda_2^2}{2\pi v_F}
\end{equation}
It learns, that $\lambda_2$ scales to strong couplings, at the standard Kondo energy scale: $T_k^{(2)}\simeq E_oe^{-2\pi v_F/J_K}$, where $E_o$ is the bare bandwidth cut-off. To obtain the precise recursion relation of the term $\lambda_3$, we consider a given static configuration of the impurity, characterized by the c-numbers $\{\tau^a\}=\pm 1/2$ (a=x,y,z). Then, the
partition function's invariance under the cut-off transformation $a=1\rightarrow a'=e^{dln L}$ simply imposes that $\lambda_3^2(a')=\lambda_3^2(a=1)(\frac{1}{a'})^{(1-K_{\rho})}$; it gives the more relevant equation:
\begin{equation}
\label{beta}
\beta(\lambda_3)=\frac{d\lambda_3}{dln L}=\frac{1}{2}(1-K_{\rho})\lambda_3=\frac{U\lambda_3}{2\pi v_F}
\end{equation}
Finally, $\lambda_3$ is well expected to become strong, at the temperature
scale\cite{sept,huit}:
\begin{equation}
T_k^{(3)}\simeq E_o.(\frac{J_K}{U})^{2\pi v_F/U}>>T_k^{(2)}
\end{equation}
We distinguish well two different regimes. When $U\rightarrow 0$, the physics is ruled by the forward Kondo sattering term, giving rise to a heavy-fermion behavior\cite{neuf}. Conversely, when $U>>J_K$, we may check that the Kondo temperature has a power-law dependence on the coupling exchange $J_K$, but also that the quenching of the impurity moment occurs for {\it ferromagnetic} as well as antiferromagnetic backward Kondo scattering $\lambda_3$. $T_k^{(3)}$ does not change under
$\lambda_3\rightarrow -\lambda_3$. The most remarkable feature in the
case of a ferromagnetic exchange $J_K$ is that a bound singlet state
is formed at the impurity site, whatever the small given Ising anisotropy of type: $\left|\Delta_z\right|\propto(\lambda_{2\perp}-\lambda_{2z})<<1$; indeed, the energy scale to create a spin S=1 in the Luttinger liquid would be:
\begin{equation}
T_{S=1}\simeq  E_o.e^{-constant/\sqrt{\left|\Delta_z\right|}}<<T_k^{(3)}
\end{equation}
It is finally due to the special nature of the spinons with S=1/2, which {\it
cannot} accept the emergence of a spin S=1 in the TL liquid. It supports, that
the presence of the Kondo backward scattering term accounts for the fundamental difference between a Kondo effect in a TL liquid and that in a Fermi
liquid. Unlike the ref.\cite{huit}, the present formalism shows explicitly, that the recursion relations of $\lambda_2$ and $\lambda_3$ are really independent up to the orders $J_K^2$ and $UJ_K$. In the weak-coupling regime $(T>>T_k^{(3)})$, the thermodynamic properties of the TL liquid, which vary as $L/v_F$ (where $L$ is the size of the sample) are not affected by the presence of the located impurity $\vec{\tau}$. However, since the term 
$\lambda_3$ changes the direction of propagation of the particles, it certainly modifies the universal conductance of the TL liquid $G_o=2e^2K_{\rho}/h$, obtained by applying a static field over a finite part of the sample. Whereas a key quantity for the Kondo effect in 3D is the resistivity, it is the conductance in 1D. In the weak-coupling regime, the small coefficient of reflection imposed by the impurity is 
defined as ${\cal R}\sim\lambda_3^2=G_o-G(T)$. Using Eq.(\ref{beta}), it suggests the following thermal correction: $G_o-G(T)\sim T^{K_{\rho}-1}$.

{\it Strong coupling limit}: As in the usual Kondo effect ($U\rightarrow 0$), we have to consider the effect of all {\it irrelevant} operators. They might be different because here the $q=0$ spin density is not involved in the strong coupling description. We work in the infinite length limit and re-consider the conductance, specific heat and 
susceptibility. Naively, the formation of a spin singlet {\bf at} the 
impurity site separates the TL liquid into two semi-infinite pieces. Unlike in Fermi
liquids, it can produce exotic tunnelling phenomena in TL ones \cite{six,huit}(cf Fig 2a). The {\bf first} irrelevant operator would come from the process in which an electron tunnels from one TL liquid to the other, with virtually breaking the spin singlet. The appropriate tunnelling is of the form
\begin{eqnarray}
\delta H_{barrier}&=&t_o\sum_{\alpha}[c^{\dag}_{1\alpha}(x=0)c_{2\alpha}(x=0)+\ (h.c)]\\ \nonumber
&=& t_o\cos[\sqrt{2\pi}\tilde{\Phi}_c(0)].trg(0)
\end{eqnarray}
where $c_1(c_2)$ is the electron operator in the left (right) semi-infinite TL liquid. Here, $t_o\sim 1/J_K^*$ denotes the bare tunnelling amplitude.  
\\
\begin{figure}[ht]
\centerline{\epsfig{file=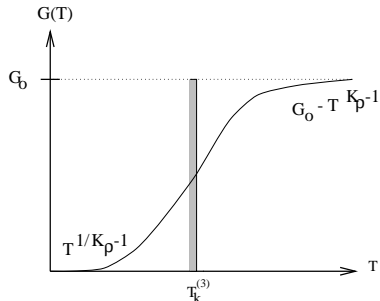,angle=0.0,height=4cm,width=5cm}}
  \caption{Behavior of the conductance in the presence of the impurity 
moment.}
 \end{figure}

The operators have to satisfy the {\it fixed boundary condition} $c_i(x=0)=0$, different from the periodic ones we have used before. This can be achieved
by fixing $\Phi_{c}(x=0)=\sqrt{\pi/2}$\cite{onze}, and by using the irrelevant
{\it dual} charge field $\tilde{\Phi}_c=\frac{1}{2}
[\tilde{\Phi}_{c1}-\tilde{\Phi}_{c2}]$ and the spin matrix $g=\frac{1}{2}[g_1-
g_2]$. Then, $t_o$ obeys a flow equation of type
of Eq.(\ref{beta}), but now replacing $K_{\rho}\rightarrow 1/K_{\rho}$. At
zero temperature, it vanishes; this is a reflection of the suppressed
density of states in a Luttinger liquid. Since the tunnelling perturbation
$t_o$ is {\it irrelevant}, the conductance varies as $G(T)\sim t_o^2(T)\sim
t_o^2T^{(1/K_{\rho})-1}$. More generally, we can consider the contribution to 
the impurity free energy of ${\cal O}(t_o^2)$. A simple rescaling argument 
implies that $f_{imp}\propto T^{1+2\Delta}$, where $(\Delta+1)$ is the 
dimension of the irrelevant 
operator. Explicitly, $\Delta=1/2(1/K_{\rho}-1)$, and we may 
confirm\cite{huit}, that it provokes deviations from the linear behavior of 
the specific heat, $C_{imp}^{(1)}\propto T^{2\Delta}\sim 
T^{(1/K_{\rho})-1}$. The susceptibility is not corrected by processes 
in ${\cal O}(t_o^2)$, since the impurity remains completely screened.  

But now, unlike ref.\cite{huit}, we argue that the Kondo effect in a TL 
liquid should generate other more crucial irrelevant operators than 
$t_o^*\rightarrow 0$. The impurity may be screened by any electron at a distance $x\leq\xi_{imp}\sim v_F/T_k^{(3)}$ of 
it (cf Fig 2b), and $\xi_{imp}$ is considerably larger than $a=1$. Since we study the physics at a length $L=1/\left|k-k_F\right|>>\xi_{imp}$, the physics in
the screening cloud can be still modeled by $\delta(x)$-interactions. Then, since $t_o^*\rightarrow 0$ or equivalently 
$\langle\cos[\sqrt{2\pi}\Phi_c(0)]\rangle\not=0$, the $q=2k_F$ fluctuations diverge around the impurity site when $T\rightarrow 0$. $\xi_{imp}$ has not the same meaning as in a Fermi liquid: here, it occurs as the {\it pinning-length} of the $q=2k_F$ spin density wave $O_{SDW}=[c^{\dag}_{L\uparrow}c_{R\downarrow}]$ 
by the impurity moment. 
\\
\begin{figure}[ht]
\centerline{\epsfig{file=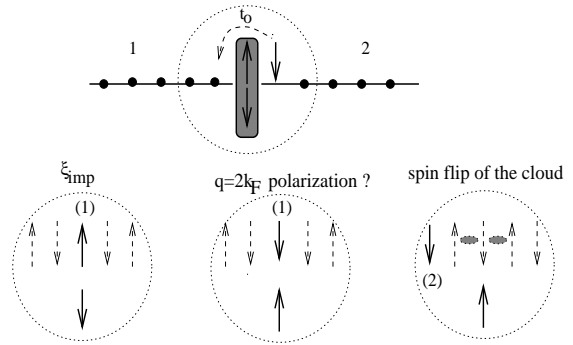,angle=0.0,height=4.5cm,width=7.3cm}}
\vskip 0.2cm 
\caption{(a) The formation of a spin singlet might lead to a possible 
tunnelling process; but $t_o$ vanishes at the fixed point. (b) Then, the 
$q=2k_F$ spin fluctuations coupled to the impurity diverge in the screening 
cloud, that engenders strong correlations between an electron (1) which 
flips the impurity spin and its two neighbors.}
\end{figure}

In fact, the low-energy description in terms of spinons, should 
definitely fail in the screening cloud due to prominent spin flips with the 
impurity. In the original Kondo effect, the impurity spin
changes the bare free electrons into free heavy-fermions. In TL liquids, we
argue that it would change weakly repulsive electrons into heavy
pseudo-particles, giving another generic non-Fermi liquid behavior.

What are the other irrelevant operators allowed by symmetry? Unlike in 
Fermi liquids\cite{neuf}, we can remark that no electron-electron interaction 
in $\eta=1/T_k^{(3)}$ may be induced by \underline{two} respective 
{\it spin flips} with the impurity. Indeed, it gives the charge term 
$\eta_1 c_{L\uparrow}^{\dag}c_{R\downarrow}c_{R\downarrow}^{\dag}c_{L\uparrow}(x=0)=-\eta_1 J_{c,L}J_{c,R}\delta(x)$, and the spin current term $-\eta_2 c_{L\uparrow}^{\dag}c_{R\uparrow}c_{R\downarrow}^{\dag}c_{L\downarrow}(x=0)=\eta_2\vec{J}_{c,L}\vec{J}_{c,R}\delta(x)$. But, $\eta_1\sim 1/U$, since $U$ defines the $q=0$-interaction energy scale in the charge sector and $\eta_2=0$ since here the $q=0$ spin density is not involved in the Kondo effect. 

{\it Typical two-channel Kondo operator}: But, if we still parallel the
well-known irrelevant operators of the standard Kondo effect in a Fermi
liquid, we have to add the following term 
\begin{eqnarray}
\delta H_{2channel}&=&\eta_3\frac{d}{dx}[c^{\dag}_{L\alpha}(x=0)c_{R\alpha}(x=0)]\\ \nonumber
&=&\eta_3\frac{d}{dx}trg(x=0)\qquad \eta_3\sim 1/\sqrt{T_k^{(3)}}
\end{eqnarray}
which respects site Parity and which is obtained through a 
$1/T_k^{(3)}$-expansion. In fact, such an operator which has the scaling 
dimension 3/2 is known to be very important in a periodic Heisenberg chain by 
varying two adjacent sites by the same amount\cite{onze}. In the present 
case, when an electron (1) comes from flipping the impurity spin, it makes all 
the spin environment unstable because disturbs the $q=2k_F$ magnetic order 
(cf Fig 2b). {\it Simultaneously}, the other electrons in the screening cloud should also flip their spins in view to restore the $q=2k_F$ antiferromagnetic polarization, strongly coupled to the impurity spin: these new collective 
spin excitations are not spinons. This should generate the expected {\it strong} antiferromagnetic links defined by $\eta_3\sim 1/{\sqrt{T_k^{(3)}}}$ between 
the electron (1) and its two neighbors. In fact, this leading irrelevant 
operator can also be written as 
$(\vec{J}_{L,-1}+\vec{J}_{R,-1}).trg\vec{\sigma}$\cite{onze} ($\vec{J}_{L,-1}$ is the 
first Kac-Moody descendent\cite{neuf}), and rewriting the theory only with 
left movers, this becomes equivalent to a single k=2 WZW field $\vec{J}_{-1}$ 
with a spin-1 primary field $\vec{\phi}$: 
$\eta_3\vec{J}_{-1}.\vec{\phi}$\cite{onzei}. We realize that $\eta_3$ is the same irrelevant operator as in the usual two-channel Kondo 
model\cite{neuf,douze}, and from second-order perturbation theory, it leads:
\begin{equation}
C_{imp}^{(2)}\propto\frac{T}{T_k^{(3)}}ln\frac{T}{T_k^{(3)}},\qquad \chi_{imp}^{(2)}\propto ln\frac{T}{T_k^{(3)}}
\end{equation}
However, the Wilson ration $R_{W}$ cannot be universal because here, there are two irrelevant operators in the theory. The low-energy behavior
is nearly the same as for the two-channel Kondo model in a free electron gas.

{\bf Mott U-transition}: Umklapps are responsible for the Mott U-transition at 
half-filling\cite{treize}. If we define incommensurability q such as
$2(k_F+q)=\pi$, the charge sector becomes massive for $q\rightarrow q_c\sim\pi/2L_c$ with the soliton length $L_c\sim\frac{1}{\sqrt{8\pi}}\sqrt{\pi v_F/U}$. It gives:
\begin{equation}
\frac{dg_3}{dln L}=2(1-K_{\rho})g_3
\end{equation}
The resulting charge gap is of the form $\Delta_c\propto 
E_o e^{-\pi v_F/U}$. Now, there is an energy gap $T\sim\Delta_c$ for creation 
of kinks, which carry S=0 and Q=2e. Then, for $T<<\Delta_c$, we must carefully 
replace $\langle\cos[\sqrt{2\pi}\Phi_c(x)]\rangle\sim constant$ and the 
renormalized exponent $K_{\rho}^*$ tends to 1/2. Since Umklapps enhance 
considerably the $q=2k_F$ spin density wave, we obtain:
\begin{equation}
\frac{d \lambda_3}{d ln L}=\frac{1}{2}(1-\frac{1}{2})\lambda_3
 \end{equation}
In that context, the Kondo temperature is \underline{rescaled} to $T_k^{(3)}
\propto E_o(J_K/U)^4<<\Delta_c$. It may be related to the problem of a spin 
defect in a spin-1/2 Heisenberg chain. But, $\xi_{imp}\sim a=1$ due to the particular value of the bare interaction $U\rightarrow +\infty$ and then, the screening cloud has no meaning\cite{onze}. Here, $\xi_{imp}$ is still much
larger than $a$: quantum fluctuations are reduced but subsist. The {\it single} permitted irrelevant operator is 
$\eta_3$, and we obtain a Wilson ratio $R_W=4/3(1+u_{\rho}^*/v_F)\rightarrow 4/3$.

Summarizing, a magnetic impurity has a great influence on TL liquids ruled by $U>J_K$. Although the impurity spin is screened, irrelevant
electronic interactions may change the collective density spin excitations of the TL liquid into heavy ``pseudo-particles'', giving the same generic non-Fermi liquid behavior as the two-channel Kondo model in a Fermi liquid.

\end{document}